\documentclass{phb-proc4-auth}

\usepackage{graphicx}
\usepackage{amssymb}

\begin{document}
\begin{frontmatter}


\journal{SCES '04}


\title{Possible canted antiferromagnetism in
UCu${}_9$Sn${}_4$}

%
%
%
%
%
%

\author[CPM]{U. Killer}
\author[CPM]{E.-W. Scheidt}
\author[Wien]{H. Michor}
\author[EPV]{J. Hemberger}
\author[EPV]{H.A. Krug von Nidda}
\author[CPM]{W. Scherer}

%

\address[CPM]{Chemische Physik und Materialwissenschaften,
Universit\"{a}t Augsburg, 86159 Augsburg, Germany}
\address[Wien]{Institut f\"ur Festk\"orperphysik, TU Wien, 1040 Wien,
Austria}
\address[EPV]{EP V, Elektronische Korrelationen und Magnetismus, Institut f\"ur Physik,
Universit\"{a}t Augsburg, 86159 Augsburg, Germany}
%
%
%
%

\thanks{This work was supported by the SFB~484 of the Deutsche
Forschungsgemeinschaft (DFG).}

%
%
%
%



\begin{abstract}
We report on the new compound UCu${}_9$Sn${}_4$ which crystallizes
in the tetragonal structure \emph{I}4/\emph{mcm} with lattice
parameters $a = 8.600{\rm\AA}$ and $c = 12.359{\rm\AA}$. This
compound is isotyp to the ferromagnetic systems RECu${}_9$Sn${}_4$
(RE = Ce, Pr, Nd) with Curie temperatures $T{}\rm_C$ = 5.5\,K,
10.5\,K and 15\,K, respectively. UCu${}_9$Sn${}_4$ exhibits an
uncommon magnetic behavior resulting in three different electronic
phase transitions. Below 105\,K the sample undergoes a valence
transition accompanied by an entropy change of 0.5~Rln2. At 32\,K
a small hump in the specific heat and a flattening out in the
susceptibility curve probably indicate the onset of helical spin
order. To lower temperatures a second transition to
antiferromagnetic ordering occurs which develops a small
ferromagnetic contribution on lowering the temperature further.
These results are strongly hinting for canted antiferromagnetism
in UCu${}_9$Sn${}_4$.
\end{abstract}

%
%



\end{frontmatter}

%
%
%
%
%

Recent studies on ternary compounds like CeNi$_9$Ge$_4$
\cite{Michor_04} and RECu${}_9$Sn${}_4$ (RE = Ce, Pr, Nd)
\cite{Singh} have attracted much interest because of their
manifold electronic and magnetic properties. In the non-Fermi
liquid (nFl) system CeNi$_9$Ge$_4$ we observed the largest ever
recorded value of the electronic specific heat $\Delta c/T
\approx$ 5.5\,J\,$\rm K^{-2}mol^{-1}$ without presence of any
trace of magnetic order \cite{Michor_04} that is mainly based on
single ion effects \cite{Killer_04}. Long range ferromagnetic
order was found in RECu${}_9$Sn${}_4$ (RE = Ce, Pr, Nd) with Curie
temperatures $T{}\rm_C$ of 5.5\,K, 10.5\,K and 15\,K, respectively
\cite{Singh}. These systems crystallize in the tetragonal
structure  \emph{I}4/\emph{mcm} with large RE-RE spacings of about
6.1\,$\rm\AA$ while the corresponding Ce-Ce distances in
CeNi$_9$Ge$_4$ value at 5.548\,$\rm\AA$. In this work we focus on
the new isotyp compound UCu${}_9$Sn${}_4$ which is on the
borderline between long range magnetic order and nFl behavior due
to the expanded 5f-electron orbital´s.
\begin{figure}
\centering
\includegraphics[width=7cm,clip]{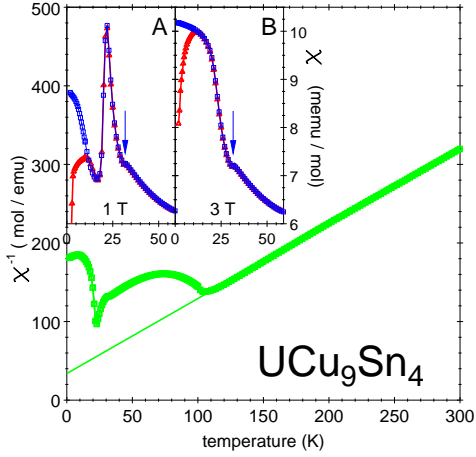}
\caption{Inverse magnetic dc susceptibility of UCu${}_9$Sn${}_4$
versus temperature measured in a magnetic field of 0.5\,T.The
straight line represents a Curie-Weiss fit to the data above
105\,K. The insert shows \emph{dc}-$\chi$ zero-field (ZFC) / field
cooled (FC) splitting in A) $B = 1$\,T and B) $B = 3$\,T}
\label{fig1}                          
\end{figure}
The X-ray powder pattern of annealed (650$^{\circ}$C for one week)
polycrystalline UCu${}_9$Sn${}_4$ also adopts the tetragonal
structure (\emph{I}4/\emph{mcm}) with $a = 8.600 \rm\AA$ and $c =
12.359 \rm\AA$ which are comparable to the lattice parameters of
RECu${}_9$Sn${}_4$. The \emph{dc}-susceptibility of
UCu${}_9$Sn${}_4$ exhibits two pronounced transitions at 105K and
22K (Fig.~\ref{fig1}). Above 105\,K the susceptibility follows a
Curie-Weiss law with $\mu$$\rm_{eff}$=2.9$\mu$$\rm_{B}$ which is
in between the f$^{1}$ and f$^{2}$ state of Uranium. The upper
transition at 105\,K, which is independent of magnetic field,
gives rise to a peak in specific heat (Fig.~\ref{fig2}). A little
anomaly is observed in resistivity measurements as well. The
entropy of this anomaly (Fig.~\ref{fig2}B) is half of that of an
antiferromagnetic transition with effective spin of 1/2 ($\Delta S
=0.5 R\ln2$) indicating a valence transition from the mixed
f$^{1}$/f$^{2}$ to f$^{1}$ state. This is confirmed by the
magnetic entropy of $S= R\ln4$ which is the same as for Ce
4f$^{1}$ in CeNi${}_9$Ge${}_4$. The expected moment of
2.54$\mu$$\rm_{B}$ is masked by strong magnetic fluctuations below
60\,K (see Fig.~\ref{fig2}A).
\begin{figure}
\centering
\includegraphics[width=6.6cm,clip]{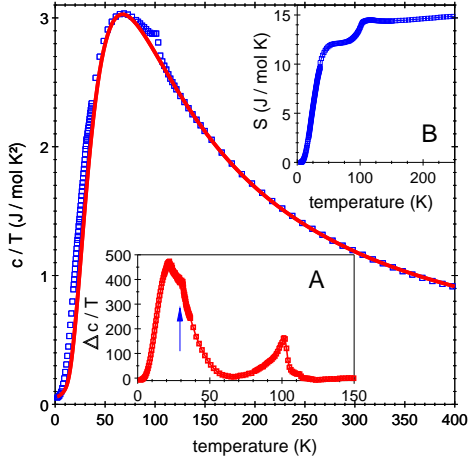}
\caption{The specific heat divided by temperature vs. temperature
of UCu$_{9}$Sn$_{4}$. The solid line is a calculation of the
phonon contribution using a Debye and two Einstein terms. A) The
electronic contribution to the specific heat $\Delta c/T$. Notice
the 3 anomalies at 22\,K, 32\,K and 105\,K. B) The entropy of the
magnetic contribution $S = R \ln4$ per U-mol at 50\,K indicates
two low lying doublets also found in CeNi${}_9$Ge${}_4$.}
\label{fig2}                          
\end{figure}

\begin{figure}
\centering
\includegraphics[width=6.5cm,clip]{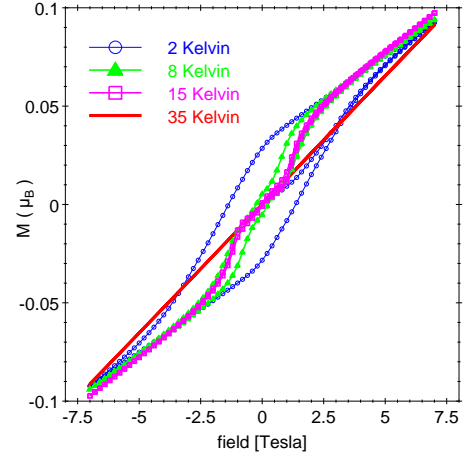}
\caption{Metamagnetic behavior and the formation of the canted
antiferromagnetic phase below 8\,K of UCu${}_9$Sn${}_4$.}
\label{fig3}                          
\end{figure}

Below 32\,K the antiferromagnetic fluctuation tends to long range
magnetic order visualized by forming out a shoulder in the
susceptibility (arrows in Fig.~\ref{fig1}A/B and
Fig.~\ref{fig2}A). The significant feature in the specific heat
and the magnetic field independent transition indicate a
helimagnetic phase \cite{Hemberger}. On decreasing temperature the
sharp peak in the \emph{dc}-$\chi$ is characteristic for a second
transition into a commensurate antiferromagnetic structure. This
is also confirmed by phonon adjusted $\Delta c/T$ data
(Fig.~\ref{fig2}A) where spin-reorientation is observed at 22\,K.

 With increasing magnetic field a slight ferromagnetic contribution develops
at low temperature indicating a sign of a canted anitferromagnetic
state. At high fields ($\geq$3\,T) a transition from the
paramagnetic straight into the canted state occurs already at
32\,K (Fig.~\ref{fig1}B). In Fig.~\ref{fig3} the magnetic
hysteresis corroborates the crossover form pure to canted
antiferromagnetism with a spin-flop transition at 15\,K and the
opening of a hysteresis loop below 8\,K with a spontaneous moment
of $M_{\rm sat}\approx$0.03$\mu$$\rm_{B}$.

The ZFC/FC splitting in \emph{dc}-$\rm\chi$ (Fig.~\ref{fig1}A/B)
may be explained by forming spontaneous small ferromagnetic
domains in the canted antiferromagnetic phase below 8\,K. In this
temperature range the ferromagnetic domains are growing in an
external magnetic field. On the occasion of the strong spin-orbit
coupling of uranium atoms the domain walls exhibit a large
inertia. The absence of any frequency shift of the transition
maximum in \emph{ac}-$\chi$ measurements rules out frustration
effects.

Summarizing three distinguished phase transitions are observed in
UCu${}_9$Sn${}_4$: i) valence shift at 105\,K, ii) onset of
helimagnetic order at 32\,K, and iii) transition into commensurate
antiferromagnetic structure at 22\,K with magnetic field induced
canting.
%
%
%
%

%
%
%
%


\end{document}